# All-electrical creation and control of giant spin-galvanic effect in 1T′-MoTe$_2$/graphene heterostructures at room temperature

Anamul Md. Hoque[1], Dmitrii Khokhriakov[1], Bogdan Karpiak[1], Saroj P. Dash[1*]

[1]Department of Microtechnology and Nanoscience, Chalmers University of Technology, SE-41296, Göteborg, Sweden

**Abstract:**

The ability to engineer new states of matter and to control their electronic and spintronic properties by electric fields is at the heart of the modern information technology and driving force behind recent advances in van der Waals (vdW) heterostructures of two-dimensional materials. Here, we exploit a proximity-induced Rashba-Edelstein (REE) effect in vdW heterostructures of Weyl semimetal candidate MoTe$_2$ and CVD graphene, where an unprecedented gate-controlled switching of spin-galvanic effect emerges due to an efficient spin-to-charge conversion at room temperature. The magnitude of the measured spin-galvanic signal is found to be an order of magnitude larger than the other systems, giving rise to a giant REE. The magnitude and the sign of the spin-galvanic signal are shown to be strongly modulated by gate electric field near the charge neutrality point, which can be understood considering the spin textures of the Rashba spin-orbit coupling-induced spin-splitting in conduction and valence bands of the heterostructure. These findings open opportunities for utilization of gate-controlled switching of spin-galvanic effects in spintronic memory and logic technologies and possibilities for realization of new states of matter with novel spin textures in vdW heterostructures with gate-tunable functionalities.

**Corresponding author:** saroj.dash@chalmers.se

The spin-orbit interaction (SOI) is the fundamental physical phenomenon and pivotal for state-of-the-art spintronics and quantum technologies[1–4]. Specifically, the SOI is the origin of the fascinating effects like current-induced transverse spin polarization in non-magnetic materials, known as spin Hall effect (SHE) in the bulk, Rashba-Edelstein effect (REE) at the heterostructure interfaces, and spin-momentum locking (SML) in topological materials[5]. Recent experiments utilizing the charge-to-spin conversion and its inverse effects have been performed on metallic multilayers[6], semiconductors[2], oxide heterostructures[7,8], two-dimensional (2D) materials[9,10], van der Waals (vdW) heterostructures with graphene[11–13], and the topological insulators[14,15]. Such SOI-induced charge-spin conversion features are promising for all-electrical spin-orbit torque-based technology[1–4].

Recently discovered topological Weyl semimetals (WSMs) were predicted to provide a much larger charge-spin conversion efficiency, the giant REE[16] and SHE[17], due to their non-trivial band structure in both bulk and surface states. The type-II WSMs such as $WTe_2$ and $MoTe_2$ possess strong SOI and band structure with a large Berry curvature and spin-polarized bulk and surface states up to room temperature[18,19]. Furthermore, the charge-spin conversion in WSMs is of great interest on its own[12,20–22], but their true potential lies in the possibilities to form vdW heterostructures with graphene. It has been predicted that addition of a graphene layer on topological materials can even enhance the REE by orders of magnitude[23–25], which can be due to different proximity-induced interactions in the heterostructures, such as Kane-Mele SOI, spin-valley coupling and Bychkov-Rashba interaction[23,26]. Moreover, an interesting energy-dependent spin texture and charge-spin conversion efficiency are also expected to be present in WSM-graphene heterostructures, like graphene-semiconductor systems[26]. The much-needed experimental proof of such giant gate tunable charge-spin conversion effect in graphene-WSM vdW heterostructure would considerably boost the chance of utilizing WSMs in emerging spintronic technologies.

Here, we report the creation of giant proximity-induced spin-galvanic effect (SGE) in vdW heterostructures of semimetal $MoTe_2$ and graphene, where a unique gate-controlled spin-charge conversion is achieved at room temperature. We employ a proximity-driven SOI mechanism, the Rashba-Edelstein effect, in the vdW heterostructure-based spintronic device to achieve spin-to-charge conversion with unprecedented efficiency. The magnitude and the sign of the spin-galvanic signal can be efficiently controlled by the application of gate voltage near the charge neutrality point, which can be understood based on the SOI-induced spin-split electronic band structure of the graphene/$MoTe_2$ heterostructure. These results highlight the unique possibility to achieve efficient gate tunability of the SGE in vdW heterostructure.

We have chosen $MoTe_2$ as the high SOI material, which has gained growing interest regarding its topological properties in the $T_d$ phase. The $MoTe_2$ is verified as the first type-II Weyl semimetal (WSM) by ARPES experiments, where tilted Weyl cones exist in pairs as the contact point between electron and hole pockets which are connected by a spin-polarized Fermi arc surface states[19,27,28]. Our basic temperature dependence measurements (Fig. S1) together with the Raman spectrum (Fig. S2) of $MoTe_2$ show the metallic 1T′ phase of the material at room temperature[29]. Because of the ability to create designer heterostructure of various 2D layered materials, we prepare the heterostructure of multilayer $MoTe_2$ of thickness 43nm (Fig. S3) and monolayer CVD graphene[30] by dry transfer method inside a glovebox.

To detect the gate-tunable spin-galvanic effect, the hybrid spintronic devices were nanofabricated in Hall-bar shaped geometry with multiple contacts as shown in the schematics and the device picture (Fig. 1a and 1b). The device consists of Hall-bar shaped graphene/MoTe$_2$ heterostructure region for spin-to-charge conversion, graphene channel for spin transport, Co/TiO$_2$ ferromagnetic contacts (FM) on graphene for spin injection, and Ti/Au contacts for detection of voltage signal and also as reference electrodes, and Si/SiO$_2$ substrate acts as a global back gate. The resistances of FM contacts on graphene are about 5-8 kΩ and the channel mobility of graphene/MoTe$_2$ heterostructure is estimated to be about 2200 cm$^2$V$^{-1}$s$^{-1}$ at room temperature.

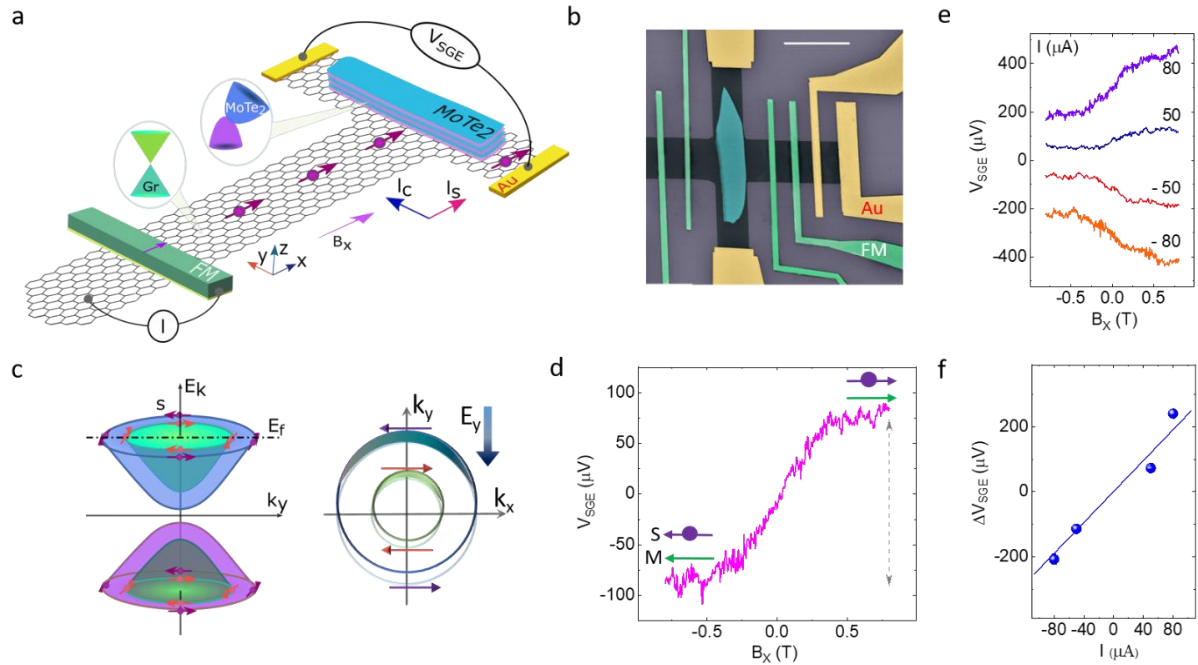

**Figure 1. Giant proximity-induced spin-galvanic effect in graphene/MoTe$_2$ van der Waals heterostructure at room temperature. (a,b)** The spintronic device schematics with the measurement scheme and scanning electron microscopic (SEM) image (false color) consisting of graphene (gray) - MoTe$_2$ (blue) heterostructure on Si/SiO$_2$ substrate, ferromagnetic (FM) and non-magnetic (Ti/Au) contacts on graphene. The scale bar in SEM image is 6 µm. **(c)** A schematic band diagram of modified graphene in a heterostructure with MoTe$_2$, showing Rashba spin-split conduction and valence bands. Application of an electric field with such spin texture is expected to create a spin accumulation due to Rashba-Edelstein effect (REE), and inversely a spin accumulation can be converted to a charge voltage (IREE). **(d)** Measured SGE as a non-local voltage ($V_{SGE}$) signal due to IREE by injecting a spin current ($I_s$) from a FM into the graphene/MoTe$_2$ heterostructure region. A change in $V_{SGE}$ is measured by sweeping a magnetic field along the x-axis ($B_x$) with an application of I=80µA, gate voltage $V_g$ = -10V for Dev 1 at room temperature. **(e,f)** Measured bias dependence of $V_{SGE}$ signal and it's magnitude $\Delta V_{SGE}$ as a function of bias current in Dev 1. A linear background is subtracted from the data and shifted in Y-axis for clarity. The line is the linear fit to the data in f.

The basic spin transport measurements were carried out at room temperature using FM injector and detector tunnel contacts in the pristine graphene ($L_{ch}$ = 5.7µm) and graphene/MoTe$_2$ heterostructure regions ($L_{ch}$ = 6.7µm), as depicted in Fig. S4. The spin transport in pristine graphene shows 200 mΩ of spin-valve signal and corresponding Hanle measurements provide an estimated spin lifetime of ~ 185 ps; on the other hand, no spin transport signal could be observed through the graphene/MoTe$_2$ heterostructure region. This disappearance of the spin signal indicates the presence of a strong SOI and band structure hybridization in the heterostructure, which is suitable for observation of spin-charge conversion phenomena. As predicted for graphene-transition metal dichalcogenides (TMD) heterostructure[26], graphene in proximity to MoTe$_2$ can also acquire strong SOI and a spin texture with a Rashba spin-split in conduction and valence bands with the same spin chirality (Fig. 1c), which can be probed by tuning the Fermi level ($E_f$) using a gate voltage.

The charge-spin conversion due to the proximityinduced SOI in graphene/$MoTe_2$ heterostructure can be measured by employing a direct REE or its inverse phenomenon (IREE) based methods[5]. The charge current-induced REE can be detected by a FM contact due to charge-to-spin conversion. In contrast, for the IREE measurements, a spin current is injected from a FM into a heterostructure region and, consequently, a voltage signal is measured due to a spin-to-charge conversion. The REE measurements can be influenced by the stray Hall effect[31] at the FM detector contact, which can mimic the SGE. Therefore, we focus on IREE method that avoids these spurious effects as non-magnetic contacts are used for detection of a non-local (NL) voltage signal. Fig. 1a shows the schematics of the NL measurement geometry used for the detection of SGE. In this hybrid spintronic device, spin current is injected from a FM contact into the graphene channel and diffused into the graphene/$MoTe_2$ heterostructure, and finally detected as a voltage signal ($V_{SGE}$) across the Hall-bar structure with non-magnetic Ti/Au contacts on graphene. Due to the IREE in the graphene/$MoTe_2$ heterostructure region with a possibility of spin-split bands, the spin density piled up on one side across the heterostructure depends on the spin direction which is detected as a voltage signal, $V_{SGE}$ ($I_c$) $\propto$ z x $n_s$ ($I_s$); here, $I_c$, z and $n_s$ are the induced charge current, out-of-plane direction and gradient of spin density accumulated via spin current $I_s$, respectively. The direction of the injected spin (in ±x-direction) and hence the $V_{SGE}$ is controlled by sweeping a magnetic field in x-direction (±$B_x$), which induces spin precession together with rotation and saturates the magnetization of the injector FM (Fig. 1d). Accordingly, the measured $V_{SGE}$ increases in the low field range and saturates at magnetic fields above B ≈ ±0.4 T (see also Fig. S5 where the x-Hanle measurements show the saturation field of the FM contact). A giant spin-galvanic signal has been obtained at room temperature with an amplitude up to $\Delta R_{SGE}$ = $\Delta V_{SGE}/I$ ~ 4.96 Ω (with I = -80μA at $V_g$ = 40V) and lower bound of the efficiency ($\alpha_{RE}$) of IREE is estimated to be 7.6% by using the equation (1)[11].

$$\alpha_{RE} = \frac{\Delta R_{SGE}}{2\rho_G} \frac{w_{MoTe_2}}{P\lambda_G(e^{-\frac{L}{\lambda_G}} - e^{-\frac{L+w_{MoTe_2}}{\lambda_G}})} \quad (1)$$

Here, $\Delta R_{SGE}$, P, $w_{MoTe_2}$(2.3 μm), $\lambda_G$ (2.65 μm), $\rho_G$ (1.6 kΩ), and L (3.37 μm) are the signal amplitude, the spin polarization of Co/$TiO_2$ contact (considering 10%), the width of the $MoTe_2$ flake, spin diffusion length, channel resistivity, and channel length, respectively. Here, we consider spin diffusion length of pristine graphene region since no spin precession signal is observed in the graphene/$MoTe_2$ heterostructure region. However, we would like to mention that the estimated $\alpha_{RE}$ is the lower limit as the spin diffusion length in graphene/$MoTe_2$ heterostructure is shorter due to the enhanced SOI. This magnitude of the observed $R_{SGE}$ and $\alpha_{RE}$ is found to be larger than the previous results on 2D materials heterostructures[11,13].

We can also electrically tune the magnitude and sign of the SGE signal with the spin injection bias current I (Fig. 1e, f). In our measurement configuration, changing the bias current on the FM spin injector contact changes the direction of accumulated spin polarization in the graphene channel. Hence, with a change in the non-equilibrium spin density, the direction of the measured $V_{SGE}$ that is induced by IREE also changes. To be noted, the bias current +I correspond to spin injection from the FM into the graphene channel and -I correspond to the spin extraction condition. The bias dependence of measured $V_{SGE}$ in the range of I = -80μA to 80μA show nearly a linear behavior with the current magnitude and changes sign by reversing the current polarity (Fig. 1f).

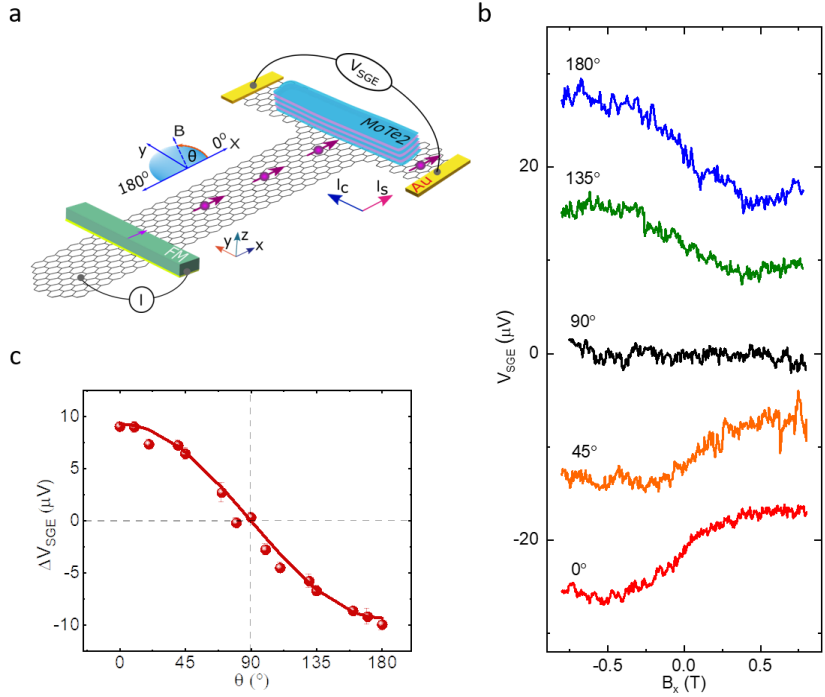

**Figure 2**. **Angle dependence of spin-galvanic effect in the graphene/MoTe$_2$ heterostructure. (a)** A schematic illustration of the SGE measurement geometry with directions of an applied magnetic field (B$_x$), the injector ferromagnet magnetization (M), the spin current (I$_s$), and external magnetic field angle (θ). **(b)** The V$_{SGE}$ measured for various measurement angle orientations, with an application of I = 60 µA for Dev 2 at room temperature. A linear background is subtracted from the data and shifted in Y-axis for clarity. **(c)** The magnitude of spin to charge conversion signals (ΔV$_{SGE}$) as a function of the applied magnetic field angle (θ). The solid line is the cos(θ) function.

Next, to substantiate the spin-origin of the experimental signal, the angular dependence of the measured NL voltage signal was performed at room temperature. The measurements were manifested by varying the angle (θ) of the applied in-plane (x-y plane) magnetic field (B) direction with respect to the x-axis, illustrated by the schematics and measurement geometry in Fig. 2a. The measured V$_{SGE}$ changes with the effective component of B in the x-direction as depicted in Fig. 2b for Dev 2 (L$_{ch}$ = 5 µm) with an injection bias current of I =60 µA. No spin-galvanic signal is detected when the B is aligned with the y-axis (θ = 90°) because at this stage, the injected spins remain parallel to B field and no spin precession is taking place in the measurement geometry. The sign of the V$_{SGE}$ reverses gradually in between θ = 0° to 180° (π) due to changing the FM magnetization direction along (±x axis) and associated reversal of the injected spin orientation. The magnitude of the measured SGE signals (ΔV$_{SGE}$) as a function of the measurement angle θ is shown in Fig. 2c, following the anticipated behavior, where the induced voltage V$_{SGE}$ (I$_c$) ∝ z x n$_s$ (I$_s$)[12,32]. The angle dependence data of the ΔV$_{SGE}$ is found to follow the cos(θ) function, which establishes the relation between the direction of the injected spin-current and the induced SGE in the graphene/MoTe$_2$ heterostructure.

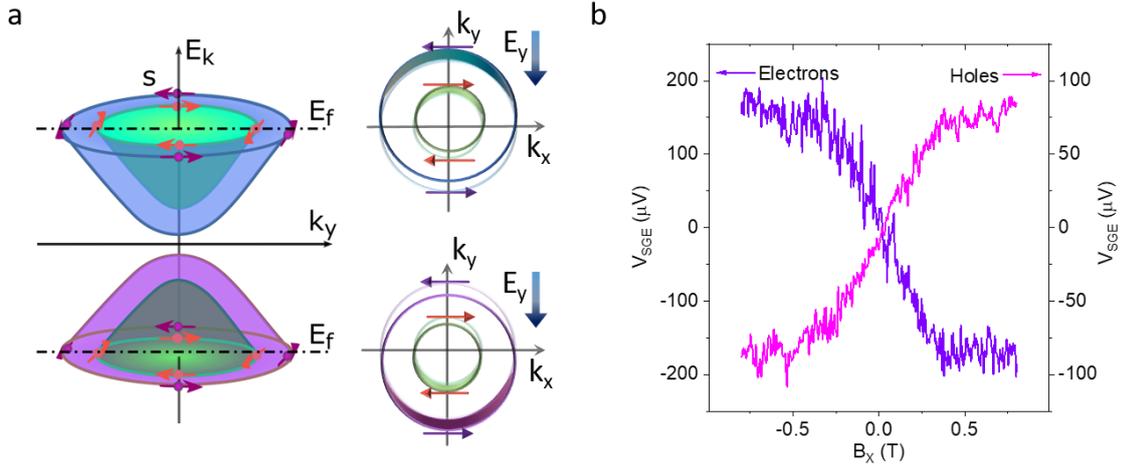

**Figure 3. Gate controlled switching of a spin-galvanic effect in graphene/MoTe$_2$ heterostructure at room temperature. (a)** Schematics of the band structure of graphene due to proximity induced SOI from MoTe$_2$. The current induced Rashba-Edelstein effect induces a non-equilibrium spin polarization depending on the Fermi-level position (E$_f$), in the valence (hole doping) and conduction (electron doping) regimes. **(b)** Measured spin-galvanic signal (V$_{SGE}$) at room temperature in the graphene/MoTe$_2$ heterostructure (Dev 1) with I = 80 µA and applied gate voltages of V$_g$= 20V and -10V to position the E$_f$ in the electron and hole doping regimes, respectively. A linear background is subtracted from the data.

In order to demonstrate the electric field-controlled switching of the SGE, we measured SGE for the electron (n) and hole (p) doped regimes of the graphene/MoTe$_2$ heterostructure. The applied gate voltage modulates the Fermi level (E$_f$) position in the graphene/MoTe$_2$ hybrid bands for probing the spin-split REE-induced non-equilibrium spin accumulation (n$_s$) with different charge carrier types, as depicted in Fig. 3(a). Fig. 3(b) shows the opposite signs of V$_{SGE}$ signals for two different doping regimes, electrons (violet) at V$_g$ = 20V and holes (magenta) at V$_g$= -10V at room temperature. Control experiments with gate dependence of standard spin-valve and Hanle spin precession signals (Fig. S6) confirm that the sign change in SGE is not due to a change in the sign of spin polarization of the injector FM contacts. The proximity-induced REE effect in graphene is predicted to be dependent on the conduction charge carrier type, i.e. electrons and holes, since spin polarity in Rashba contour remains unchanged in valence and conduction bands, unlike the spin-momentum locking in topological insulators [26,33]. The observed IREE produces a charge current (I$_c$ ∝ z x n$_s$) perpendicular to the injected spin densities (n$_s$), which changes its direction because of a unidirectional accumulation of electron and holes (-q for electrons and +q for holes) in the heterostructure with specific spin orientations. So, it can be concluded that the measured SGE signal and the sign change of the V$_{SGE}$ is inherently due to induced IREE in graphene/MoTe$_2$ heterostructure. Similar gate-dependent sign change behavior of SGE is also observed in Dev 2 (see Fig. S7).

Systematic gate dependence of V$_{SGE}$ is measured at different gate voltages (V$_g$) in the range from -40 to +40V for two different spin injection bias currents I = +/- 80µA at room temperature (Fig. 4a). It can be observed that the SGE signal changes sign with gate voltage and spin injection bias currents. To correlate the sign change behavior of SGE signal with the carrier type of the graphene/MoTe$_2$ heterostructure, the gate dependence of graphene channel resistance (R$_{Gr}$) is plotted together with ΔV$_{SGE}$ and ΔR$_{SGE}$. To be noted, the R$_{Gr}$ is measured for the channel length of 6.7 µm (Fig. 4d inset), which includes part of the graphene region that is not contributing to the SGE signal. Therefore, the charge neutrality point (CNP) for the only graphene/MoTe$_2$ heterostructure region can be between 10-20 V. Interestingly, the gate dependence of SGE signal ΔV$_{SGE}$ and ΔR$_{SGE}$ change sign around the charge neutrality point, which is between 10-20 V. Opposite signs of the SGE signals are consistently observed for carriers in the electron- and hole-doped regimes of the heterostructure, i.e. for the conduction and valence bands of the hybrid structure. The signal amplitudes $\Delta V_{SGE}$ and $\Delta R_{SGE}$ are also found to be asymmetric, higher for electron-doped regime compared to the hole-doped regime. Such behavior can be due to the different doping regimes and spin-charge conversion efficiency.

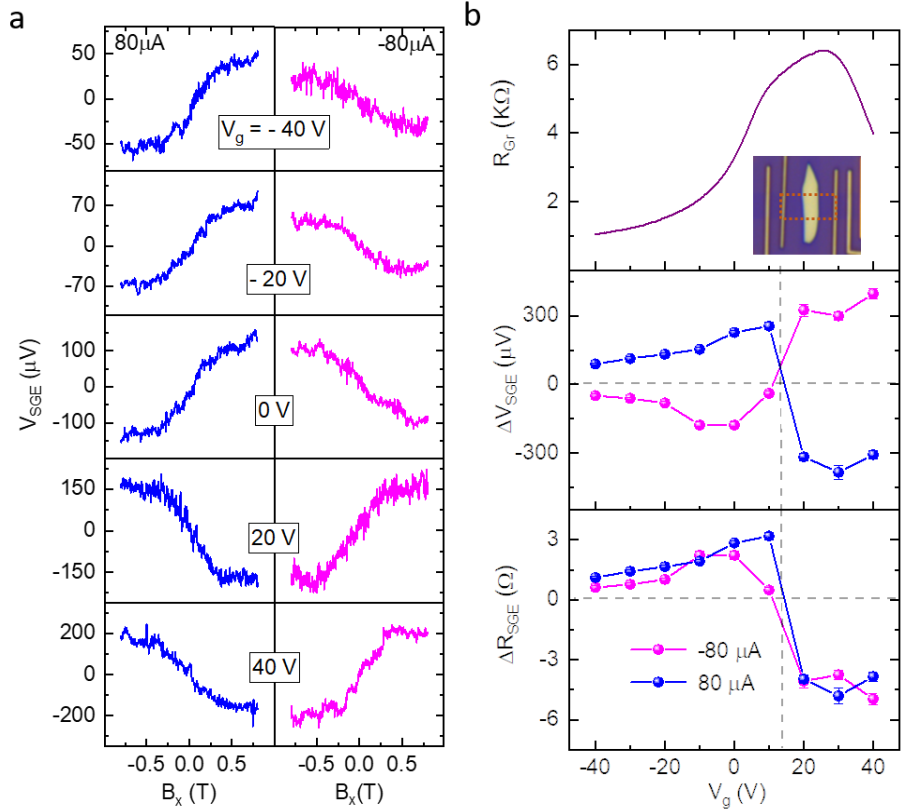

**Figure 4. Gate dependence of spin-galvanic effect in graphene/MoTe$_2$ heterostructure at room temperature. (a)** Evolution of measured spin-galvanic signal V$_{SGE}$ for different gate voltages in the range of +40 to -40V, with I = +/- 80μA for Dev 1 at room temperature. The sign of the spin-galvanic signal changes with gate voltage V$_g$ and spin injection bias current I. **(b)** Gate dependence of graphene channel resistance (R$_{Gr}$) and spin-galvanic signals ($\Delta V_{SGE}$ and $\Delta R_{SGE} = \Delta V_{SGE}/I$) at I = +/- 80μA at room temperature. The sign change of the spin-galvanic signal is observed close to the charge neutrality point of graphene and the signals are of opposite sign for n and p-type regimes of the graphene/MoTe$_2$ heterostructure.

The observed giant and gate-tunable SGE in graphene/MoTe$_2$ heterostructure can have fascinating microscopic origins. Although topologically non-trivial Weyl states in MoTe$_2$ are found to exist at low temperature[27], the trivial spin-polarized surface and bulk states are also shown to exist in these materials even at room temperature[19]. Additionally, the existence of polar instability near the surface is also observed even when the MoTe$_2$ bulk is in a centrosymmetric phase[19,27]. Therefore, the surface states of MoTe$_2$ can have a strong influence on the band hybridization and proximity-induced spin-charge conversion properties of graphene/MoTe$_2$ heterostructure. Additionally, we expect the presence of a very strong SOI with band splitting of few meV in graphene in proximity with MoTe$_2$, similar to the predictions on graphene/semiconducting TMD heterostructure[23,34]. Interestingly, the induced Rashba spin-split bands in graphene should have opposite spin polarization with a helical spin texture due to the lack of inversion symmetry at the interface, which is protected by time-reversal symmetry[4,26]. Moreover, this proximity-induced band splitting in graphene is predicted to emanate along the energy axis with an in-plane spin polarization, which is the cardinal aspect for a novel spin-to-charge conversion effects[34]. The in-plane spin accumulation in our measured graphene/$MoTe_2$ heterostructures renders this SGE signal to be due to proximity-induced IREE in graphene. It can be affirmed that the polarization of the accumulated spins are in-plane, otherwise, a signal would have been observed when applied magnetic field direction is aligned with the y-axis (θ = 90°) in our angle dependence measurements[11,13]. We do not expect the SHE and REE from the $MoTe_2$ surface states to contribute to the manifested gate-tunable SGE signals as the charge carriers in metallic $MoTe_2$ are not tunable by a gate voltage. We can also rule out proximity-induced SHE in graphene in a heterostructure with $MoTe_2$, since that would have polarized the spins in z-direction, which would have resulted in a spin precession Hanle peaks near B = 0T field, which are not observed[13]. In contrast, we observe a gate dependence tuning of the spin signal magnitude and a sign change for the electron- and hole-doped regimes of the graphene/$MoTe_2$ heterostructure. So, the observed SGE is inevitably due to IREE in graphene because of proximity-induced SOI of $MoTe_2$.

Note: After the preparation of this manuscript, we became aware of the recent preprint on graphene/TMD heterostructure[35,36]. Here we use semimetal 1T' $MoTe_2$/graphene heterostructure, which results in the observation of a giant IREE effect, orders of magnitude larger than previous reports. The explicit demonstration of the gate tunability and switching of the SGE at room temperature and the use of large-area CVD graphene in our devices emphasize the importance for future applications.

In conclusion, we demonstrated the electrical detection and control of proximity-induced Rashba spin-orbit interaction in vdW Heterostructures at room temperature. The observed giant SGE in heterostructure of 1T′ semimetallic $MoTe_2$ and large-area CVD graphene can offer unique advantages for faster and low-power spin-charge conversion based spintronic devices. Specifically, vdW heterostructure provides entirely new mechanisms for tuning emergent spin functionalities by an electric field, resulting from interfacial proximity interactions between the stacked layered materials. Due to proximity effect, graphene acquires strong SOI and a spin texture with spin-split conduction and valence bands, which is ideal for tuning by a gate voltage. Importantly, controlling of the magnitude and sign of the spin voltage by an external electric field at room temperature can be efficiently used for gate tunable spin-orbit torque (SOT) based magnetic random-access memory (MRAM) [3] and logic technologies[2]. The strong Rashba SOI in vdW heterostructures can further enable new energy-dependent spin textures by tuning their proximity-induced interactions and expected to provide a new playground for creation of unique topological states of matter, for example, in heterostructures with layered ferromagnetic and superconducting materials[1,4].

## Methods

The vdW heterostructures of graphene/MoTe$_2$ are fabricated by dry transfer of 1T′ MoTe$_2$ (from HQ Graphene) flakes on CVD graphene (from Groltex), onto SiO$_2$ (285 nm)/n-doped Si substrate using scotch tape method inside a glovebox in an inert atmosphere. The graphene and the graphene/MoTe$_2$ channels are patterned into Hall-bar shaped structures by electron beam lithography and oxygen plasma etching. Both non-magnetic and ferromagnetic contacts to graphene were defined by two-step electron beam lithography and lift-off process. The non-magnetic contacts were prepared by deposition of 10 nm Ti and 60 nm Au by electron beam evaporation and lift-off process. For the preparation of ferromagnetic tunnel contacts to graphene, a two-step electron beam evaporation of 0.6 nm of Ti and in-situ oxidation, followed by a 90 nm of Cobalt (Co) deposition. The lift-off was performed in warm acetone and IPA. The spin-galvanic measurements are performed in a vacuum cryostat with a magnetic field up to 0.8 Tesla and sample rotation stage. All the measurements are performed at room temperature using a current source Keithley 6221, a nanovoltmeter Keithley 2182A and Keithley 2612B source meter for application of gate voltages.

## Data availability

The data that support the findings of this study are available from the corresponding authors on reasonable request.

## Affiliations

[1]Department of Microtechnology and Nanoscience, Chalmers University of Technology, SE-41296, Göteborg, Sweden

## Acknowledgments

The authors acknowledge financial support from the EU FlagEra project (from Swedish Research Council VR No. 2015-06813), Swedish Research Council VR project grants (No. 2016-03658), European Union's Horizon 2020 research and innovation program under grant agreements no. 696656 and no. 785219 (Graphene Flagship Core 1 and Core 2), Graphene center and the EI Nano and AoA Materials Science program at Chalmers University of Technology. We acknowledge the help and discussions with Bing Zhao in our group and staff at Quantum Device Physics and Nanofabrication laboratory in our department at the Chalmers University of Technology.

## Contributions

SPD and AMH conceived the idea and designed the experiments. AMH fabricated and measured the devices. DK, BK, SPD contributed in some device fabrication and measurements. AMH and SPD analyzed the experimental data and wrote the manuscript. SPD supervised the research and managed the research project.

## Competing interests

The authors declare no competing financial interests.

## Corresponding authors:

Correspondence and requests for materials should be addressed to Saroj P. Dash, Email: saroj.dash@chalmers.se